\documentclass{JHEP3}
\usepackage{multicol,bbm}
\usepackage{graphicx,psfrag}
\usepackage{dcolumn}
\usepackage{bm}
\usepackage{amsmath,amssymb}
\newcommand{\eq}{\begin{equation}}
\newcommand{\eqe}{\end{equation}}

\newcommand{\eqa}{\begin{eqnarray}}
\newcommand{\eqae}{\end{eqnarray}}

\JHEPspecialurl{http://jhep.sissa.it/JOURNAL/JHEP3.tar.gz}

\title{Novel aspects of C-theories in Cosmology}

\author{Tomi S. Koivisto\\
Institute of Theoretical Astrophysics, University of Oslo,
0315 Oslo, Norway}
\author{David F. Mota\\
Institute of Theoretical Astrophysics, University of Oslo,
0315 Oslo, Norway}
\author{Marit Sandstad\\
Institute of Theoretical Astrophysics, University of Oslo,
0315 Oslo, Norway}

\received{\today} %%
\accepted{} %% These are for published papers.
\abstract{The field equations in FRW background for the so called C-theories are presented and investigated. In these theories the usual Ricci scalar is substituted with $f(\mathcal{R})$ where $\mathcal{R}$ is a Ricci scalar related to a conformally scaled metric $\hat{g}_{\mu\nu} = \mathcal{C}(\mathcal{R})g_{\mu\nu}$, where the conformal factor itself depends on $\mathcal{R}$. It is shown that homogeneous perturbations of this Ricci scalar around general relativity FRW background of a large class of these theories are either inconsistent or unstable.} 
\preprint{}
\keywords{palatini, modified gravity, cosmology, C-theories}

\begin{document}

  \section{Introduction}
  When the first supernovae results \cite{Riess_et_al1998,Perlmutter_et_al1999} were published in the late nineties, it was official: The universe seemed to be dominated by something that was giving it an accelerated expansion. The most recent observational data from supernova \cite{Conley_et_al2011,Suzuki_et_al2012}, baryonic accoustic oscillations \cite{Percival_et_al2010,Padmanabhan_et_al2012,Beutler_et_al2011,Blake_et_al2011,Anderson_et_al2012} and CMB measurements \cite{Bennett_et_al2012,PlanckXVI2013} provide strong and consistent support for a $\Lambda$CDM concordance model of cosmology, which provides the mathematically simplest explanation for this.

  However, the $\Lambda$CDM has problems. From a theoretical point of view the cosmological constant \cite{Weinberg1989,Martin2012} poses a serious challenge to reconcile its magnitude with naturalness arguments. If it should have come from a particle physics vacuum energy, its value is about 55 orders of magnitude too small \cite{Martin2012}. Observational anomalies have also been detected for instance in the CMB data \cite{Bennett2010,PlanckXXIII2013}, though the severity of these issues are much smaller.

  These problems combined with scientific curiosity have lead to a wide variety of alternative theories that attempt to explain the observational results. The most obvious way of changing the model is perhaps to use a form of dynamical dark energy \cite{Wetterich1987,Ratra1987,Caldwell1998} instead of a cosmological constant. The number of such theories are vast as can be seen from recent reviews on the subject \cite{Copeland2006,Frieman2008,Li2011,Polarski2012,Tsujikawa2013}.

  A second way of responding to these problems could be  to consider the assumptions that lead to the postulation of a cosmological constant or dark energy component. One possibility could be to question the hypothesis of a homogenous and isotropic underlying geometry, which is at least wrong at sufficiently small scales, see for instance \cite{Andersson_and_Coley2011} and references therein. Another approach would be to modify the theory of general relativity on scales where it has not been vigourously tested. In recent years the study of such modified gravities and their effects on cosmology has attracted considerable attention, for recent reviews see \cite{Clifton_et_al2012,Nojiri_et_Odintsov2011,Capozziello_et_Laurentis2011}. For a slightly older review, but which includes comparison to dark energy models see \cite{Durrer2008}. \cite{Yoo2012} is a recent review which goes through all the above mentioned alternatives explaining the accelerated expansion.

  Many of the alternative approaches to the accelerated expansion have been ruled out after their introduction to the community. Thorough examination of stability and cosmological viability of any new proposal is necessary in order to see if the new model should be excluded or if more data is necessary to break the degeneracy between it and other models in explaining observables. This paper is an attempt at making a small contribution to this process for one particular class of modified gravity models. 

  \cite{Amendola_Enqvist_and_Koivisto2011} introduced a general modified gravity framework called the C-theories which unifies the Einstein and Palatini gravities\footnote{ For a recent review on the Palatini variational principle see \cite{Olmo2011}. For some of its applications to modified gravities and physical  implications see \cite{Koivisto:2005yc,pal1,pal2,Koivisto:2007sq,pal3,pal4,Koivisto:2010jj}. Other alternative variational principles for gravity can be be described within a generalised version of this framework as was shown in \cite{Koivisto032011}. For another type of interpolation between Palatini and metric theories, see \cite{Harko:2011nh,Koivisto:2013kwa}.}  In addition to introducing a wide class of completely new theories, it contains a smooth transition of theories between the Einstein and Palatini variational principles for $f(R)$ gravities.\footnote{For a brief review on f(R) and related modified gravity theories see \cite{Bergliaffa2011}.} The post Newtonian (PPN) parameters for the theory were calculated in \cite{Koivisto092011} and application of a similar theory to inflation was considered in \cite{Enqvist_Koivisto_and_Rigopoulos2011}. Applications to cosmological evolution in the late universe was made in \cite{Finke2012}. In the recent paper \cite{Sandstad_et_al2013} it was shown that a certain subset of these theories are equivalent with a certain class of conformal non-local gravity theories first described in \cite{Biswas_Mazumdar_and_Siegel,Gottlober_et_al1990,Schmidt1990}.

In this paper we present the field equations for this theory in an FRW background. We reformulate these equations in a way useful for further investigation. We use this new formulation to study linearised versions of classes of sub-theories close to GR and show that many of these are unstable or inconsistent close to GR. 
%;-)

The paper is divided into four sections following this introduction. Section \ref{sec:Definition} gives an introduction to the theory. In section \ref{sec:FieldEquations} we give the field equations in general and in particular on an FRW background. The equations are also reformulated in order to facilitate numerical work or linearised study. Section \ref{sec:Linearisation} holds a study of the linearised version of certain subclasses of the theory in FRW and finds them to be inconsistent or unstable for use in cosmology. It is also shown that this inconsistency naturally vanishes when viewing traditional $f(R)$ theories in the C-theory framework. In section \ref{sec:Conclusions} the results are reviewed, and implications for future work are discussed.

\section{The C-theory defined}\label{sec:Definition}
The model introduces a conformal factor $\mathcal{C}$ which is a function of the Ricci curvature scalar $\mathcal{R}$. That is we have two metrics one unhatted and one hatted, conformally related in the following way:
  \begin{equation}\label{eq:hattedUnhattedConformalRelation}
    \hat{g}_{\mu\nu} = \mathcal{C}(\mathcal{R})g_{\mu\nu}
  \end{equation}
  where:
  \begin{equation}
    \mathcal{R} = g^{\mu\nu} \hat{R}_{\mu\nu}
  \end{equation}
  $\hat{R}_{\mu\nu}$ is the normal Ricci tensor in the hatted metric. Using equation (\ref{eq:hattedUnhattedConformalRelation}) we can get expressions for the hatted quantities in terms of the unhatted ones and $\mathcal{C}$:\footnote{There was a typo in the equation for the Christoffel symbol corresponding to equation \ref{eq:ChristoffelHat} in the original article \cite{Amendola_Enqvist_and_Koivisto2011}.}
  \begin{eqnarray}
    \hat{\Gamma}^\gamma_{\alpha \beta} &=& \Gamma^\gamma_{\alpha \beta} + \frac{1}{2}\left(2\delta^\gamma_{(\alpha}\delta^\lambda_{\beta)} - g_{\alpha\beta}g^{\gamma\lambda}\right)\nabla_{\lambda}\ln \mathcal{C}\label{eq:ChristoffelHat}\\
    \mathcal{R} &=& R - \frac{\left(D-1\right)}{4\mathcal{C}^2}\left[4\mathcal{C}\Box\mathcal{C} + \left(D-6\right)\left(\partial\mathcal{C}\right)^2\right]\label{eq:ReccursiveR}
  \end{eqnarray}
  where $D$ is the number of dimensions.

  The action for the theory is then taken to be:
  \begin{equation}
    S = \int d^Dx\sqrt{-g}\left[f(\mathcal{R}) + 16\pi G\mathcal{L}_m\left(\Psi, \nabla_\alpha\Psi, g_{\mu\nu}\right)\right]
  \end{equation}
  
  To get an action for this which is a bit easier to variate, we treat the two metrics as independent and use a Lagrange multiplier $\lambda^{\mu\nu}$ to impose the conformal relation (\ref{eq:hattedUnhattedConformalRelation}):
  \begin{eqnarray}
    \lambda &\equiv& g_{\mu\nu}\lambda^{\mu\nu} \qquad \hat{\lambda} \equiv \hat{g}_{\mu\nu}\lambda^{\mu\nu}\\
    S &=& \int d^Dx \sqrt{-g}\left[f\left(\mathcal{R}\right) + \hat{\lambda} - \mathcal{C}\left(\mathcal{R}\right)\lambda + 16\pi G\mathcal{L}_m\left(\Psi, \nabla_\alpha\Psi, g_{\mu\nu}\right)\right]
  \end{eqnarray}

  As was shown in the original paper \cite{Amendola_Enqvist_and_Koivisto2011}, there is also a scalar tensor formulation of this with $\mathcal{L}_g$ the Lagrangian density of the (modified) gravitational part of the action:
\begin{equation}
  \mathcal{L}_g = f(\phi) + \hat{\lambda} - \mathcal{C}\lambda + \xi\left(\mathcal{R} - \phi\right)
\end{equation}
where $\xi$ is a new Lagrangian multiplier, ensuring that $\phi = \mathcal{R}$. By using the equation of motion for $\phi$, we can solve for $\phi$ in terms of $\xi$ and $\lambda$. Variating the action with respect to $\lambda_{\mu\nu}$ allows us to get rid of the hatted metric entirely. In the end the tensor structure of $\lambda$ does not contribute and we get an action with two scalar fields $\xi$ and $\lambda$ given by the Lagrangian density:
\begin{equation}\label{eq:ScalarTensorFormulation}
  \mathcal{L}_g = \xi R + \frac{D-1}{4\mathcal{C}^2}\left[4\mathcal{C}g^{\mu\nu}\xi_{,\mu}\mathcal{C}_{,\nu} -\left(D - 2\right)\xi\left(\partial \mathcal{C}\right)^2\right] - \xi\phi + f\left(\phi\right)
  \end{equation}
  where $\phi = \phi\left(\xi,\lambda\right)$ and $\mathcal{C} = \mathcal{C}\left(\phi\left(\xi, \lambda\right)\right)$
  
  Looking more closely at equation (\ref{eq:ScalarTensorFormulation}) we realise that $\lambda$ only enters indirectly in expressions for $\phi$ upon which the function $\mathcal{C}$ is defined. This means that the dynamics are equal to what they would have been if one just considered the $\phi$ and $\xi$ fields, hence this is what we do.
 % However, since only $\xi$ is a dynamical field, the equation of motion for $\lambda$ will only give an algebraic constraint equation which we can use to eliminate $\lambda$. We are then left with a single scalar field theory, for the field $\xi$

In passing we mention that by implementing the constraint on the connections (one then needs a three-index lagrange multiplier), instead of the underlying metrics, would result in an equivalent theory. We omit the proof here. 

%\section{Relations to other modified gravity theories}

  \section{Cosmology in the two-scalar formulation}\label{sec:FieldEquations}
  We want to develop a framework in which to study the cosmology of these theories. To do this we will be working in the scalar tensor formulation with the Lagrangian density given by equation (\ref{eq:ScalarTensorFormulation}). We will also assume minimal coupling to the matter fields, that is we are working in the Jordan frame. We of course also restrict our search to four spacetime dimensions.

  To recapitulate our action becomes
  \begin{equation}
    S = \int d^4x\sqrt{-g}\left[\xi R + \frac{3g^{\mu\nu}\xi_{,\mu}\mathcal{C}_{,\nu}}{\mathcal{C}} - \frac{3}{2}\xi\frac{\left(\partial \mathcal{C}\right)^2}{\mathcal{C}^2} - \xi\phi + f(\phi)\right] + S_m
  \end{equation}
  or using the fact that $\mathcal{C} = \mathcal{C}(\phi)$ to get the derivatives of $\phi$ explicitly:
  \begin{equation}
    S = \int d^4x\sqrt{-g}\left[\xi R + 3g^{\mu\nu}\xi_{,\mu}\phi_{,\nu}\frac{\mathcal{C}'}{\mathcal{C}} - \frac{3}{2}\xi\left(\frac{\mathcal{C}'}{\mathcal{C}}\right)^2\left(\partial\phi\right)^2 - \xi\phi + f\right] + S_m
  \end{equation}
  where we from now on denote functions of $\phi$ such as $f(\phi)$ and $\mathcal{C}(\phi)$, simply as $f$ and $\mathcal{C}$.

  \subsection{Equations of motion and field equations}
  We variate the action with respect to the metric to get the field equations:
    \begin{eqnarray}
    8\pi G T_{\mu\nu} &=& \left(R_{\mu\nu} - \frac{1}{2}g_{\mu\nu}R\right)\xi + 3\frac{\mathcal{C}'}{\mathcal{C}}\left[\phi_{,\mu}\left(\xi_{,\nu} - \frac{1}{2}\frac{\mathcal{C}'}{\mathcal{C}}\xi\phi_{,\nu}\right) - \frac{1}{2}g_{\mu\nu}g^{\alpha\beta}\phi_{,\alpha}\left(\xi_{,\beta} - \frac{1}{2}\xi\frac{\mathcal{C}'}{\mathcal{C}}\phi_{,\beta}\right)\right] \nonumber \\
    &&+ \frac{1}{2}g_{\mu\nu}\left(\xi\phi - f\right) - \left(\nabla_\mu\nabla_\nu - g_{\mu\nu}\Box\right)\xi\label{eq:NonvacField}
  \end{eqnarray}
  with trace:
  \begin{equation}\label{eq:NonvacFieldTrace}
    8\pi G T^{\mu}_{\mu} = - R\xi - 3\frac{\mathcal{C}'}{\mathcal{C}}\left[g^{\alpha\beta}\phi_{,\alpha}\left(\xi_{,\beta} - \frac{1}{2}\frac{\mathcal{C}'}{\mathcal{C}}\xi\phi_{,\beta}\right)\right] + 2\left(\xi\phi - f\right) +3 \Box\xi
  \end{equation} 

  We then find the equations of motion for $\xi$ and $\phi$. Seeing as we are working in the Jordan frame, the matter components do not contribute to these. Starting with the one for $\xi$, we get:
  \begin{equation}\label{eq:EOMXI}
    R + \frac{3}{2}\frac{\left(\partial \mathcal{C}\right)^2}{\mathcal{C}^2} - 3\frac{\Box \mathcal{C}}{\mathcal{C}}  - \phi =     R + \frac{3}{2}\left(\frac{\mathcal{C}'}{\mathcal{C}}\right)^2\left(\partial\phi\right)^2 - 3\frac{\mathcal{C}''}{\mathcal{C}}\left(\partial\phi\right)^2 - 3\frac{\mathcal{C}'}{\mathcal{C}}\Box\phi - \phi = 0
  \end{equation}
  Then the one for $\phi$
  \begin{equation}\label{eq:EOMPHI}
   0 = f' - \xi - 3\frac{\mathcal{C}'}{\mathcal{C}}\Box\xi + 3g^{\mu\nu}\xi_{,\mu}\phi_{,\nu}\left(\frac{\mathcal{C}'}{\mathcal{C}}\right)^2 + 3\xi\left(\partial\phi\right)^2\frac{\mathcal{C}'}{\mathcal{C}}\left(\frac{\mathcal{C}''}{\mathcal{C}} - \left(\frac{\mathcal{C}'}{\mathcal{C}}\right)^2\right)  + 3\xi\left(\frac{\mathcal{C}'}{\mathcal{C}}\right)^2\Box\phi
   \end{equation}
   Using the trace of the field equation (\ref{eq:NonvacFieldTrace}) to get rid of the $\Box\xi$ term:
   \begin{eqnarray}
    0 &=&f' - \xi -\frac{\mathcal{C}'}{\mathcal{C}}\left\{8\pi GT_\mu^\mu + R\xi + 3\frac{\mathcal{C}'}{\mathcal{C}}\left[g^{\alpha\beta}\phi_{,\alpha}\left(\xi_{,\beta} - \frac{1}{2}\frac{\mathcal{C}'}{\mathcal{C}}\xi\phi_{,\beta}\right)\right] + 2\left(f\left(\phi\right)- \xi\phi\right)\right\}\nonumber \\
   &&+ 3g^{\mu\nu}\xi_{,\mu}\phi_{,\nu}\left(\frac{\mathcal{C}'}{\mathcal{C}}\right)^2 + 3\xi\left(\partial\phi\right)^2\frac{\mathcal{C}'}{\mathcal{C}}\left(\frac{\mathcal{C}''}{\mathcal{C}} - \left(\frac{\mathcal{C}'}{\mathcal{C}}\right)^2\right) + 3\xi\left(\frac{\mathcal{C}'}{\mathcal{C}}\right)^2\Box\phi\nonumber \\
   &=& -\xi\left(1 -\phi\frac{\mathcal{C}'}{\mathcal{C}}\right) + f' - 2\frac{\mathcal{C}'}{\mathcal{C}}\left(f + 4\pi GT_\mu^\mu\right) -\frac{\mathcal{C}'}{\mathcal{C}}\xi\left[R +\frac{3}{2}\left(\frac{C'}{\mathcal{C}}\right)^2\left(\partial\phi\right)^2 - 3\frac{\mathcal{C}''}{\mathcal{C}}\left(\partial\phi\right)^2  - 3\frac{\mathcal{C}'}{\mathcal{C}}\Box\phi  - \phi\right]\nonumber \\
  \end{eqnarray}
  and finally the equation of motion for $\xi$ (\ref{eq:EOMXI}) takes care of the last parenthesis yielding:
  \begin{equation}\label{xiSolve}
      \xi = \frac{\mathcal{C}f' - 2\mathcal{C}'\left(f + 4\pi G T^{\mu}_{\mu}\right)}{\mathcal{C} - \phi\mathcal{C}'}
  \end{equation}
  this means that we can reduce to a theory of only one field in vacuum as was shown in \cite{Koivisto092011}.

  \subsection{Equations of motion and field equations in an FRW background}
To find the impact of this theory on the cosmological history at the background level, we need to look at the equations in the homogenous and isotropic case. That is $\phi = \phi(t)$, $\xi = \xi(t)$ and we assume the Friedmann Robertson Walker metric. We also take the universe to be flat:
  \begin{equation}
    ds^2 = -dt^2 + a^2(dx^2 + dy^2 + dz^2)
  \end{equation}
  In this case the equation of motion for $\xi$ (\ref{eq:EOMXI}) becomes:
  \begin{equation}\label{eq:EOMXIHom}
    6\left(\frac{\ddot{a}}{a} + H^2\right) - \frac{3}{2}\left(\frac{\mathcal{C}'}{\mathcal{C}}\right)^2\dot{\phi}^2 + 3\frac{\mathcal{C}''}{\mathcal{C}}\dot{\phi}^2 + 3\frac{\mathcal{C}'}{\mathcal{C}}\left(\ddot{\phi} + 3H\dot{\phi}\right) - \phi = 0
  \end{equation}
  The equation of motion for $\phi$ stays the same (\ref{xiSolve}) and we want a couple of the Einstein equations. A convenient choice will be the $00$ component, giving the first Friedmann equation:
  \begin{equation}\label{eq:FriedmannHom}
    8\pi G\rho = 3H^2\xi + \frac{3}{2}\frac{\mathcal{C}'}{\mathcal{C}}\left[\dot{\phi}\dot{\xi} - \frac{1}{2}\frac{\mathcal{C}'}{\mathcal{C}}\xi\dot{\phi}^2\right] + \frac{1}{2}\left(f - \xi\phi\right) + 3H\dot{\xi}
  \end{equation}
  As our second choice we pick the Raychaudri equation given by $-\frac{1}{6}$ times the $00$ part plus the spatial trace:
  \begin{equation}\label{eq:RaychaudriHom}
    \frac{\ddot{a}}{a}\xi = -\frac{4\pi G}{3}\left(\rho + 3 P\right) + \frac{\mathcal{C}'}{\mathcal{C}}\dot{\phi}\dot{\xi} - \frac{1}{2}\left(\frac{\mathcal{C}'}{\mathcal{C}}\right)^2\xi \dot{\phi}^2 + \frac{1}{6}\left(\xi\phi - f\right) - \frac{1}{2}\ddot{\xi}- \frac{1}{2}H\dot{\xi}
  \end{equation}
  
  From this we define a $\rho_\phi$ and a $P_{\phi}$ stemming from the C-theory. We do this by interpreting equations (\ref{eq:FriedmannHom}) and (\ref{eq:RaychaudriHom}) as:
  \begin{eqnarray}
    3H^2 &=& 8\pi G \left(\rho + \rho_\phi\right) \\
    \frac{\ddot{a}}{a} &=& 8\pi G\left(-\frac{1}{6}\left(\rho + 3P\right) - \frac{1}{6}\left(\rho_\phi + 3P_{\phi}\right)\right)
  \end{eqnarray}
  We then find that:
  \begin{eqnarray}
    \rho_{\phi} &=& \frac{1}{8\pi G}\left[\frac{1}{2}\left(\xi\phi - f\right) -\frac{3}{2}\frac{\mathcal{C}'}{\mathcal{C}}\dot{\phi}\left(\dot{\xi} - \frac{1}{2}\frac{\mathcal{C}'}{\mathcal{C}}\xi\dot{\phi}\right) - 3H\dot{\xi} - 3H^2\xi + 3H^2\right]\label{eq:densityPhi}\\
    P_\phi &=& \frac{1}{8\pi G}\left[- \frac{1}{2}\left(\xi\phi - f\right) -\frac{3}{2}\frac{\mathcal{C}'}{\mathcal{C}}\dot{\phi}\left(\dot{\xi} - \frac{1}{2}\frac{\mathcal{C}'}{\mathcal{C}}\xi \dot{\phi}\right)  + \ddot{\xi} + 2H\dot{\xi}   + \left(2\dot{H} + 3H^2\right)\left(\xi - 1\right)\right]\nonumber\\\label{eq:PressurePhi}
  \end{eqnarray}

  We observe that for theories with regimes dominated by the first term (proportional to $\xi\phi - f$), $w_\phi = \frac{P_\phi}{\rho_\phi}$ will approach $-1$. For theories dominated by the second term (proportional to $\dot{\xi} - \frac{1}{2}\frac{\mathcal{C}'}{\mathcal{C}}\xi \dot{\phi}$) it will go to $w_\phi = 1$. Hence we see that there is at least potential for accelerated expansion-like solutions from this, which we of course expect since this encompasses $f(R)$-gravities.

  \subsection{Reformulated equations in e-folding time}
 The equations (\ref{eq:EOMXIHom}) to (\ref{eq:RaychaudriHom}) will not be analytically solvable in the general case, and we need to simulate the background evolution of the universe. To get a general idea of what is going on, it can be useful first to linearise the equations close to GR. To do both of these things it will be convenient to consider the equations in terms of e-folding time $\frac{d}{dt} \to H \frac{d}{d\ln a}$. The different matter components of the universe will follow their usual adiabatic expansion equations 
\begin{equation}\label{eq:efoldingAdiabaticExp}
  \dot{\rho_i} = -3\left(1 + w_i\right)\rho_i
  \end{equation}
  where $\dot{\rho_i}$ now denotes the e-folding time derivative $\frac{d}{d\ln a}$ of the density of particle species $i$. The notation $\frac{d}{d\ln a} = \dot{}$ will be used alongside the notation $\frac{d}{dt} = \dot{}$. The former for a while now, and notice will be given whenever the notation is changed.

  Since the evolutions of the energy densities is given for each value of the conformal time, given the initial conditions, there are three different quantities for which we need to set up a set of three coupled differential equations. These are $\phi$, $\dot{\phi}$ and $H$. For $\phi$ of course, the differential equation wil be completely trivial, $\dot{\phi} = \dot{\phi}$. 

  To get an equation for $\dot{H}$, we simply solve for this quantity in the e-folding time version of equation (\ref{eq:EOMXIHom}):
  \begin{equation}\label{eq:genDerivHforSim}
    \dot{H} = \frac{\frac{\phi}{3H} - H\left(4 + \frac{\mathcal{C}'}{\mathcal{C}}\left(\ddot{\phi} + 3\dot{\phi}\right) + \dot{\phi}^2\left(\frac{\mathcal{C}''}{\mathcal{C}} - \frac{1}{2}\left(\frac{\mathcal{C}'}{\mathcal{C}}\right)^2\right)\right)}{2 + \frac{\mathcal{C}'}{\mathcal{C}}\dot{\phi}}
  \end{equation}

  For the purposes of getting a cleaner starting point for a linearised analysis or a possible simulation of the equations, we need an equation for $\ddot{\phi}$ which does not depend on $\dot{H}$. One way of doing this is by equating the expressions for $6\left(\frac{\ddot{a}}{a} + H^2\right)\xi$ given in equation (\ref{eq:EOMXIHom}) with that from a combination of equations (\ref{eq:FriedmannHom}) and(\ref{eq:RaychaudriHom}), and then get rid of the remaining $\dot{H}$ (stemming from the redefinition to e-folding time) by using equation (\ref{eq:genDerivHforSim}). The first part is done working with $\frac{d}{dt} = \dot{}$:
  \begin{eqnarray}
    6\left(\frac{\ddot{a}}{a} + H^2\right)\xi &=&\frac{3}{2}\left(\frac{\mathcal{C}'}{\mathcal{C}}\right)^2\xi\dot{\phi}^2 - 3\frac{\mathcal{C}''}{\mathcal{C}}\xi\dot{\phi}^2 - 3\frac{\mathcal{C}'}{\mathcal{C}}\left(\ddot{\phi}\xi + 3H\dot{\phi}\xi\right) + \phi\xi \nonumber\\
    &=& 8\pi G\left(\rho - 3 P\right) + 3\frac{\mathcal{C}'}{\mathcal{C}}\dot{\phi}\dot{\xi} - \frac{3}{2}\left(\frac{\mathcal{C}'}{\mathcal{C}}\right)^2\xi \dot{\phi}^2 + 2\xi\phi - 2f - 3\ddot{\xi}- 9H\dot{\xi} \nonumber\\
    0 &=&-8\pi GT^{\mu}_{\mu}  + \xi\phi - 2f + 3\left(\frac{d}{dt} + 3H\right)\left(\frac{\mathcal{C'}}{\mathcal{C}}\dot{\phi}\xi - \dot{\xi}\right) \label{eq:NumSolveForddotphi1}
  \end{eqnarray}

  To get further we need the expressions for $\dot{\xi}$  and $\ddot{\xi}$:
  \begin{equation}\label{eq:dotXi}
    \dot{\xi} = \frac{1}{1 - \phi \frac{\mathcal{C}'}{\mathcal{C}}}\left[\dot{\phi}\left[f'' - \frac{\mathcal{C}'}{\mathcal{C}}f' - 2\frac{\mathcal{C}''}{\mathcal{C}}\left(f + 4\pi GT^{\mu}_{\mu}\right) + \phi\frac{\mathcal{C}''}{\mathcal{C}}\xi\right]- 8\pi G\frac{\mathcal{C}'}{\mathcal{C}}\dot{T}^{\mu}_{\mu}\right]
  \end{equation}
  \begin{eqnarray}
    \ddot{\xi} &=& \frac{1}{1 - \phi \frac{\mathcal{C}'}{\mathcal{C}}}\Bigg\{\dot{\phi}^2\Big[\frac{\phi\frac{\mathcal{C}''}{\mathcal{C}}}{1 - \phi \frac{\mathcal{C}'}{\mathcal{C}}}\left(f'' - \frac{\mathcal{C}'}{\mathcal{C}}f' - 2\frac{\mathcal{C}''}{\mathcal{C}}\left(f + 4\pi GT^{\mu}_{\mu}\right) + \phi\frac{\mathcal{C}''}{\mathcal{C}}\xi\right) + \nonumber\\
    &&f''' - \frac{\mathcal{C}'}{\mathcal{C}}f'' - 2\frac{\mathcal{C}'''}{\mathcal{C}}\left(f + 4\pi G T^{\mu}_{\mu}\right)- 2\frac{\mathcal{C}''}{\mathcal{C}}f' + \frac{\mathcal{C}''}{\mathcal{C}}\xi + \phi\frac{\mathcal{C}'''}{\mathcal{C}}\xi\Big] +  \nonumber\\
    &&\ddot{\phi}\left[f'' - \frac{\mathcal{C}'}{\mathcal{C}}f' - 2\frac{\mathcal{C}''}{\mathcal{C}}\left(f + 4\pi GT^{\mu}_{\mu}\right) + \phi\frac{\mathcal{C}''}{\mathcal{C}}\xi\right]+ \nonumber\\
    &&\dot{\phi}\left[\phi\frac{\mathcal{C}''}{\mathcal{C}}\dot{\xi} - 16\pi G\frac{\mathcal{C}''}{\mathcal{C}}\dot{T}^{\mu}_{\mu} - \frac{8\pi G\dot{T}^{\mu}_{\mu}\frac{\mathcal{C}'}{\mathcal{C}}\phi\frac{\mathcal{C}''}{\mathcal{C}}}{1 - \phi \frac{\mathcal{C}'}{\mathcal{C}}}\right] - 8\pi G\frac{\mathcal{C}'}{\mathcal{C}}\ddot{T}^{\mu}_{\mu}\Bigg\}\label{eq:ddotXi}
  \end{eqnarray}
  We insert this and the expression for $\xi$ (\ref{xiSolve}) into equation (\ref{eq:NumSolveForddotphi1}) and change to e-folding time derivatives ($\dot{} = \frac{d}{d\ln a}$). The result we obtain solving for $\ddot{\phi}$ is:
\begin{equation}\label{eq:NumSolveForddotphiFinal}
  \ddot{\phi} = \frac{1}{A}\left(B\dot{\phi}^3 + C\dot{\phi}^2 + D\dot{\phi} + E\right)
\end{equation}
where:
\begin{eqnarray}
  A &=& -f'' + \left(2\frac{\mathcal{C}'}{\mathcal{C}} - \frac{\frac{\mathcal{C}''}{\mathcal{C}}\phi}{1 - \phi\frac{\mathcal{C}'}{\mathcal{C}}}\right)f' + 2\left(\frac{\frac{\mathcal{C}''}{\mathcal{C}}}{1 - \phi\frac{\mathcal{C}'}{\mathcal{C}}} - \left(\frac{\mathcal{C}'}{\mathcal{C}}\right)^2\right)\left(f +  4\pi G T^{\mu}_{\mu}\right) - 4\left(\frac{\mathcal{C}'}{\mathcal{C}}\right)^2\pi G\dot{T}^{\mu}_{\mu}\nonumber\\
  B &=& \frac{1}{2}\frac{\mathcal{C}'}{\mathcal{C}}f''' - \left[\frac{1}{2}\frac{\mathcal{C}''}{\mathcal{C}} + \frac{1}{4}\left(\frac{\mathcal{C}'}{\mathcal{C}}\right)^2 - \frac{\frac{\mathcal{C}''}{\mathcal{C}}\phi\frac{\mathcal{C}'}{\mathcal{C}}}{1 - \phi\frac{\mathcal{C}'}{\mathcal{C}}}\right]f''  + \nonumber\\
  &&\left[\left(\frac{\mathcal{C}'}{\mathcal{C}}\right)^3 - \frac{\frac{\mathcal{C}''}{\mathcal{C}}}{1 - \phi\frac{\mathcal{C}'}{\mathcal{C}}}\left(\frac{\mathcal{C}'}{\mathcal{C}} + \frac{\mathcal{C}''}{\mathcal{C}}\phi + \frac{1}{2}\left(\frac{\mathcal{C}'}{\mathcal{C}}\right)^2\phi -2\frac{\frac{\mathcal{C}''}{\mathcal{C}}\frac{\mathcal{C}'}{\mathcal{C}}\phi^2}{1 - \phi\frac{\mathcal{C}'}{\mathcal{C}}}\right) + \frac{\frac{\mathcal{C}'''\mathcal{C}'}{\mathcal{C}^2}\phi}{1 - \phi\frac{\mathcal{C}'}{\mathcal{C}}}\right]\frac{f'}{2} \nonumber\\
  &&+\left[\left( \frac{\mathcal{C}''}{\mathcal{C}}\right)^2-\frac{1}{2}\left(\frac{\mathcal{C}'}{\mathcal{C}}\right)^4 - \frac{\frac{\mathcal{C}'''\mathcal{C}'}{\mathcal{C}^2}}{1 - \phi\frac{\mathcal{C}'}{\mathcal{C}}} - \frac{\frac{\mathcal{C}''\mathcal{C}'}{\mathcal{C}^2}}{1 - \phi\frac{\mathcal{C}'}{\mathcal{C}}}\left(\frac{1}{2}\frac{\mathcal{C}'}{\mathcal{C}} + \frac{\mathcal{C}''}{\mathcal{C}}\phi + 2\frac{\frac{\mathcal{C}''}{\mathcal{C}}\frac{\mathcal{C}'}{\mathcal{C}}\phi^2}{1 - \phi\frac{\mathcal{C}'}{\mathcal{C}}}\right)\right]\left(f + 4\pi GT^{\mu}_{\mu}\right)\nonumber\\
  C &=& f'''  - \frac{\mathcal{C}'}{\mathcal{C}}f''+ \frac{\mathcal{C}''}{\mathcal{C}}\phi\frac{2f''}{1 - \phi\frac{\mathcal{C}'}{\mathcal{C}}}- \frac{\mathcal{C}''}{\mathcal{C}}f' + 2\left(\frac{\mathcal{C}'}{\mathcal{C}}\right)^2f'\nonumber\\
  && -\left(\frac{2\mathcal{C}''}{\mathcal{C}} - \frac{\mathcal{C}'''}{\mathcal{C}}\phi - 2\frac{\left(\frac{\mathcal{C}''}{\mathcal{C}}\right)^2\phi^2}{1 -\phi\frac{\mathcal{C}'}{\mathcal{C}}}\right)\frac{f'}{1 - \phi\frac{\mathcal{C}'}{\mathcal{C}}}  - 2\left(-\frac{\mathcal{C}''\mathcal{C}'}{\mathcal{C}^2}+ \left(\frac{\mathcal{C}'}{\mathcal{C}}\right)^3\right)\left(f + 4\pi GT^{\mu}_{\mu}\right)\nonumber\\
  &&-2\left(\frac{\mathcal{C}'''}{\mathcal{C}} + 2\frac{\left(\frac{\mathcal{C}''}{\mathcal{C}}\right)^2\phi}{1 -\phi\frac{\mathcal{C}'}{\mathcal{C}}}\right)\frac{f + 4\pi G T^{\mu}_{\mu}}{1 - \phi\frac{\mathcal{C}'}{\mathcal{C}}} - \frac{\mathcal{C}'}{\mathcal{C}}4\pi G\dot{T}^{\mu}_{\mu}\left(\frac{\mathcal{C}''}{\mathcal{C}} - \frac{1}{2}\left(\frac{\mathcal{C}'}{\mathcal{C}}\right)^2 + 2\frac{\frac{\mathcal{C}''}{\mathcal{C}}\phi\frac{\mathcal{C}'}{\mathcal{C}}}{1 - \phi\frac{\mathcal{C}'}{\mathcal{C}}}\right)\nonumber\\
  D &=& \frac{\phi}{3H^2}\left[\frac{1}{2}f'' + \left(\left(\frac{\mathcal{C}'}{\mathcal{C}}\right)^2 - \frac{\mathcal{C}''}{\mathcal{C}} + \frac{\mathcal{C}'}{\mathcal{C}\phi}\right)\left(f +  4\pi G T^{\mu}_{\mu}\right) - \frac{3}{2}\frac{\mathcal{C}'}{\mathcal{C}}f' + \frac{\mathcal{C}''}{\mathcal{C}}\phi\frac{\frac{1}{2}f' - \frac{\mathcal{C}'}{\mathcal{C}}\left(f + 4\pi G T^{\mu}_{\mu}\right)}{1 - \phi\frac{\mathcal{C}'}{\mathcal{C}}}\right]\nonumber\\
  &&+ f'' - 2\left(\frac{\mathcal{C}''}{\mathcal{C}} - \left(\frac{\mathcal{C}'}{\mathcal{C}}\right)^2\right)\left(f +  4\pi G T^{\mu}_{\mu}\right) - 2\frac{\mathcal{C}'}{\mathcal{C}}f' + \frac{\mathcal{C}''}{\mathcal{C}}\phi\frac{f' - 2\frac{\mathcal{C}'}{\mathcal{C}}\left(f + 4\pi G T^{\mu}_{\mu}\right)}{1 - \phi\frac{\mathcal{C}'}{\mathcal{C}}}\nonumber\\
  &&+8\pi G \dot{T}^{\mu}_{\mu}\left[\left(\frac{\mathcal{C'}}{\mathcal{C}}\right)^2 -2\frac{\frac{\mathcal{C''}}{\mathcal{C}}}{1 - \phi\frac{\mathcal{C}'}{\mathcal{C}}}\right] -  4\pi G\ddot{T}^{\mu}_{\mu}\left(\frac{\mathcal{C}'}{\mathcal{C}}\right)^2\nonumber\\
  E &=& \frac{1}{3H^2}\left[ 2\left(f + 4\pi GT^{\mu}_{\mu}\right)- \phi\left(f' + \frac{\mathcal{C}'}{\mathcal{C}}4\pi G\dot{T}^{\mu}_{\mu}\right)\right] - 8\frac{\mathcal{C}'}{\mathcal{C}}\pi G\left(\dot{T}^{\mu}_{\mu} +\ddot{T}^{\mu}_{\mu}\right)\label{eq:NumSolveForddotphiCoeffs}
\end{eqnarray}

It will be useful to have an expression for the energy-momentum tensor and its derivatives. It is defined as:
\begin{equation}
  T^{\mu}_{\mu} = -\rho + 3p = \sum_i \left(-\rho_i + 3p_i\right) = \sum_i \left(-1 + 3w_i\right)\rho_i  
\end{equation}
where we sum over the independent universe components $i$ with equations of state $w_i$.
From equation (\ref{eq:efoldingAdiabaticExp}) we then find that:
\begin{eqnarray}
  \dot{T}^{\mu}_{\mu} &=&  = \sum_i 3\left(1 - 3w_i\right)\left(1 + w_i\right)\rho_i \\
  \ddot{T}^{\mu}_{\mu} &=& - 9\sum_i \left(1 - 3w_i\right)\left(1 + w_i\right)^2\rho_i 
\end{eqnarray}

For a universe containing matter, radiation and a cosmological constant this becomes:
\begin{equation}\label{eq:EMTensorEfoldingDerivative}
  T^{\mu}_{\mu} = -\frac{3H_0^2}{8\pi G}\left(\Omega_m + 4\Omega_\Lambda\right) \qquad \dot{T}^{\mu}_{\mu} = \frac{9H_0^2}{8\pi G}\Omega_m \qquad \ddot{T}^{\mu}_{\mu} = -\frac{27H_0^2}{8\pi G}\Omega_m
\end{equation}
where for later convenience we have introduced the density parameters $\Omega_x = \frac{8\pi G}{3H_0^2}\rho_x$, which retain the dependencies on e-folding time of the densities, i.e. $\Omega_m \propto e^{-3x}$, $\Omega_\gamma \propto e^{-4x}$and $\Omega_\Lambda$ is constant.

 \section{Linearised equations}\label{sec:Linearisation}
 It is interesting to study whether different types of C-theories are stable around the usual general relativity. We then assume that the theory can be set very close to general relativity and linearise the equations to find stability conditions. Since the equations are quite involved we first list the definitions we have chosen for the linearised versions of the different functions:
 \begin{eqnarray}
   \phi &=& \mathcal{R} = R + \delta\phi\nonumber\\
   f\left(\phi\right) &=& f(R) + f'(R)\delta\phi\nonumber\\
   \mathcal{C}\left(\phi\right) &=& \mathcal{C}\left(R\right) + \mathcal{C}'\left(R\right)\delta\phi
 \end{eqnarray}
 where $R$ should be interpreted as the pure GR solution for the Ricci tensor.
 
 When higher derivatives of the functions $f$ and $\mathcal{C}$ are involved we must always go one order up to obtain the linearised term:
\begin{equation}
  f^{(n)}\left(\phi\right) = f^{(n)}(R) + f^{(n+1)}(R)\delta\phi
\end{equation}
etc. 
\subsection{A special case close to GR }

As the full linearised version of equation (\ref{eq:NumSolveForddotphiFinal}) turns out to be quite complicated, we will make some assumptions on $f$ and $\mathcal{C}$. One particularly simple scheme for the linearisation will then be:
 \begin{eqnarray}
   \phi &=& \mathcal{R} = R + \delta\phi\nonumber\\
   f\left(\phi\right) &=& \phi = R + \delta\phi\nonumber\\
   \frac{\mathcal{C}'\left(\phi\right)}{\mathcal{C}\left(\phi\right)} &=& \alpha
 \end{eqnarray}
where $\alpha$ is a parameter that we consider to be small and of the same order as $\delta \phi$. Since the coefficient $A$ in equation (\ref{eq:NumSolveForddotphiFinal}) is first order in this quantity, we will need to go to second order in the small parameters in the coefficients to obtain results for $\delta\ddot{\phi}$. We assume first then that the equations above are exact identities even to second order, and that:
\begin{equation}
  \frac{\mathcal{C}^{(n)}}{\mathcal{C}} = \alpha^n
\end{equation}

Then the coefficients in (\ref{eq:NumSolveForddotphiCoeffs})read
 \begin{eqnarray}
  A &=& \alpha \left[2 - \alpha R - 4\alpha\pi G\dot{T}^{\mu}_{\mu}\right] \qquad B = 0 \qquad C =  -\alpha^2 \nonumber\\
  D &=& \alpha\left[\frac{\left(8\pi G T^{\mu}_{\mu} + \alpha R^2\right)}{6H^2} - 2 + \alpha R - 8\alpha\pi G \dot{T}^{\mu}_{\mu} -  4\alpha\pi G\ddot{T}^{\mu}_{\mu}\right]\nonumber\\
  E &=& \frac{1}{3H^2}\left[\delta\phi -  4\pi \alpha G\dot{T}^{\mu}_{\mu}\left(R + \delta\phi\right)\right] - 8\alpha\pi G\left(\dot{T}^{\mu}_{\mu} +\ddot{T}^{\mu}_{\mu}\right)\label{eq:NumSolveForddotphiCoeffsLinearisedspecial}
\end{eqnarray}
where we have also used that for GR $R = -8\pi G T^\mu_\mu$. Putting all this together like in equation (\ref{eq:NumSolveForddotphiFinal}) we get this:
\begin{eqnarray}
  \ddot{R} + \delta\ddot{\phi} &=& -\frac{\alpha}{2}\dot{R}^2 + \left(\frac{4\pi G T^\mu_\mu}{3H^2} - 1 + \frac{\left(2\pi G\right)^2\alpha T^\mu_\mu\dot{T}^\mu_\mu}{3H^2} - \frac{\alpha R}{2} - 6\alpha\pi G\dot{T}^\mu_\mu - 2\alpha\pi G\ddot{T}^\mu_\mu\right)\dot{R}\nonumber\\
  && + \left(\frac{4\pi G T^\mu_\mu}{3H^2} - 1\right)\delta\dot{\phi} + \frac{1}{6H^2}\left[\frac{\delta\phi}{\alpha} -  4\pi G\dot{T}^{\mu}_{\mu}\left(R + \delta\phi\right)\right] - 4\pi G\left(\dot{T}^{\mu}_{\mu} +\ddot{T}^{\mu}_{\mu}\right)\nonumber\\
  && - \frac{1}{6H^2}\left[\delta\phi\left(\frac{R}{2} + 2\pi G\dot{T}^\mu_\mu\right) -  4\pi \alpha G\dot{T}^{\mu}_{\mu}R\left(\frac{R}{2} + 2\pi G\dot{T}^\mu_\mu\right)\right] \nonumber\\
  &&- 4\alpha\pi G\left(\dot{T}^{\mu}_{\mu} +\ddot{T}^{\mu}_{\mu}\right)\left(\frac{R}{2} + 2\pi G\dot{T}^\mu_\mu\right)
\end{eqnarray}
separating the equation, we realise that the zeroth order equation becomes an equation for $\delta\phi$:
\begin{equation}\label{eq:linearisedDeltaPhi}
  \delta\phi = \alpha\left[3H^2\left(\ddot{R} + \dot{R}\right) + \frac{R\dot{R}}{2}\right]
\end{equation}
which proves that $\delta\phi$ is order $\alpha$ small in this setup.
The first order part then yields:
\begin{eqnarray}
  \delta\ddot{\phi} &=&  \frac{\alpha}{2}\dot{R}\ddot{R}+\frac{\alpha}{2}\dot{R}^2 - \frac{\alpha}{2}R\dot{R} - \left(\frac{R}{6H^2} + 1\right)\delta\dot{\phi} + \frac{\alpha\dot{R}^2R}{16H^2}\label{linearisdeDdotDeltaPhi1}
\end{eqnarray}
We can now take the double derivative of $\delta\phi$ as given in equation (\ref{eq:linearisedDeltaPhi}) to check if we obtain consistent results. To do this we make use of the fact that in GR $\dot{H} = \frac{R}{6} - 2H^2$
\begin{eqnarray}
  \delta\dot{{\phi}} &=& \alpha\left[H\left(R - 12H^2\right)\left(\ddot{R} + \dot{R}\right) + 3H^2\left(\dddot{R} + \ddot{R}\right) + \frac{\dot{R}^2}{2} + \frac{R\ddot{R}}{2}\right]\nonumber\\
  \delta\ddot{\phi} &=& \alpha\Bigg[\frac{\left(R - 12H^2\right)^2}{6}\left(\ddot{R} + \dot{R}\right) + H\left(\dot{R} - 4HR + 48H^2\right)\left(\ddot{R} + \dot{R}\right) + \nonumber\\
    &&2H\left(R - 12H^2\right)\left(\dddot{R} + \ddot{R}\right) + 3H^2\left(\ddddot{R} + \dddot{R}\right) + \frac{3}{2}\dot{R}\ddot{R} + \frac{R\dddot{R}}{2}\Bigg]\label{linearisdeDdotDeltaPhi2}
\end{eqnarray}
equating the two expressions (\ref{linearisdeDdotDeltaPhi1}, \ref{linearisdeDdotDeltaPhi2}) for $\delta\ddot{\phi}$ we get:
\begin{eqnarray}
  0 &=& \ddot{R}\left(\dot{R} + R\right) +  \frac{R\dot{R}}{2}  + \frac{R}{12H^2}\left(\frac{\dot{R}^2}{4} + R\ddot{R}\right) - H\left(R + 12H^2\right)\dot{R}\nonumber\\
  && +\frac{R^2}{6H}\left(\ddot{R} + \dot{R}\right) + 4RH\left(\ddot{R}+ \dddot{R}\right) - 24H^3\left(2\dddot{R} + 3\ddot{R}\right)\nonumber\\
  && +\left(\ddot{R} + \dot{R}\right)\left(\frac{R^2}{6} - 8RH^2 + 72H^4 + \dot{R}H\right) \nonumber\\
  && + 3H^2\left(\ddddot{R} + 2\dddot{R} + \ddot{R}\right) + R\dddot{R}\label{eq:LinearisedConstraint}
\end{eqnarray}
As the time derivatives here are with respect to e-folding time, it is not difficult to rephrase this as a constraint on the universe content of the theory. For convenience we state the value the GR background Ricci scalar should take in terms of universe components, seeing as we assume it to be a GR solution. We also assume that the universe contains only matter, radiation and a cosmological constant:
\begin{eqnarray}
  R &=& -8\pi G T^\mu_\mu = 3H_0^2\left(\Omega_{m} + 4\Omega_\Lambda\right) \qquad \mathrm{and} \qquad H^2 = H_0^2\left(\Omega_{\gamma} + \Omega_{m} + \Omega_\lambda\right)
\end{eqnarray}
What values its e-folding derivatives take can now easily be found from equation (\ref{eq:EMTensorEfoldingDerivative}).

Before applying all these formulas to equation (\ref{eq:LinearisedConstraint}) we note that the equation will always be fulfilled if $\Omega_{m0}$ is zero. This is not a realistic late time universe model.\footnote{However, if one considered the C-theory instead to be important at very early times, this might not be problematic.} If this is not the case, the equation will give a complicated constraint on the relation between the universe components. Assuming the universe to be flat, we already have one constraint on the components, that is that they should sum to $1$ at the time of definition denoted by $0$. In a universe with no dark energy $\Omega_\Lambda = 0$, this constraint would then fully constrain the parameters. Otherwise some freedom is left, but probably not enough to satisfy constraints from late time experiments. In this case the constraint becomes:
\begin{eqnarray}
  0 &=& \Omega_m\Bigg[12\Omega_\gamma-\frac{7}{2}\Omega_m - 14\Omega_\Lambda + \frac{3\left(\Omega_m+ 4\Omega_\Lambda\right)}{4\left(\Omega_m + \Omega_\gamma + \Omega_\Lambda\right)}\left(\frac{5}{4}\Omega_m + 4\Omega_\Lambda\right) \nonumber\\
  &&+ H_0\sqrt{\Omega_m + \Omega_\gamma + \Omega_\Lambda}\left(28\Omega_\gamma + 15\Omega_m\right)+ \frac{H_0\left(\Omega_m  + 4\Omega_\Lambda\right)^2}{\sqrt{\Omega_m + \Omega_\gamma + \Omega_\Lambda}}\nonumber\\
  && +H_0^2\left(33\Omega_m^2 + 24\Omega_m\Omega_\Lambda + 80\Omega_m\Omega_\gamma + 32\Omega_\Lambda\Omega_\gamma + 48\Omega_\gamma^2\right)\Bigg]
\end{eqnarray}

We note, that though we chose a particular model in this case, the analysis bears a somewhat wider generality. As long as $f(R)$ is linear in $R$ to zeroth order in some linearisation parameter, and $\mathcal{C}(R)$ is constant in $R$ to zeroth order, none of which are entirely farfetched constraints for theories close to GR, the zeroth order of equation (\ref{eq:NumSolveForddotphiFinal}) will result in an equation for $\delta\phi$ which is independent of $\delta\dot{\phi}$ and $\delta\ddot{\phi}$, but that will contain terms of order up to $\ddot{R}$. The second derivative in time of this equation, will hence yield an equation of up to $\ddddot{R}$ for $\delta\ddot{\phi}$ which will not be trivially identical to the first order linearised version of (\ref{eq:NumSolveForddotphiFinal}). From equating the two expressions, in general a constraint on the universe content for such a C-theory to be consistent will emerge, which in generic cases will render the theory inconsistent. Hence one must look for theories further away from GR, or with functions $\mathcal{C}$ and $f$ that are far from the naive close to GR guesses to find consistent C-theories.

\subsection{The linearised palatini case}
Another set of assumptions for linearisation which is interesting to consider is the one given below:
 \begin{eqnarray}
   \phi &=& \mathcal{R} = R + \delta\phi\nonumber\\
   \mathcal{C}\left(\phi\right) &=& 1 + \alpha\phi = 1 + \alpha R\nonumber\\
   f'\left(\phi\right) &=&  \mathcal{C}\left(\phi\right) \qquad \mathrm{i.e.} \qquad f = R + \delta\phi + \frac{\alpha}{2} R^2\nonumber\\
 \end{eqnarray}
where again $\alpha$ is a parameter that we consider to be small and of the same order as $\delta \phi$. This is the linearised version of the Palatini $f(R)$ theories. We will take it that the assumption of $\mathcal{C}$ being a linear function in $\phi$ is exact, though the main points of the analysis will not change significantly if this assumption is lifted. Again our assumptions also include that $R$ is the exact GR response to the universe content, so that $R = -8\pi G T^\mu_\mu$. The coefficients in equation (\ref{eq:NumSolveForddotphiCoeffs}) then become:
\begin{eqnarray}
  A &=& \alpha\left[1 - \alpha R + \frac{\alpha}{2}\dot{R}\right] \qquad B = 0 \qquad C = \alpha^2\nonumber\\
  D &=& \alpha\left[\frac{R}{6H^2}\left(\alpha R  - 1\right) - 1 + \alpha\left(R - \dot{R} + \frac{\ddot{R}}{2}\right)\right]\nonumber\\
  E &=& \frac{1}{3H^2}\left[\delta\phi + \frac{\alpha}{2}\left(R +\delta\phi - \alpha R\right)\dot{R}\right] + \alpha\left(1 - \alpha R\right)\left(\dot{R} +\ddot{R}\right)
\end{eqnarray}
as in the previous case, we realise that the zeroth order expression is an equation for $\delta\phi$:
\begin{equation}\label{eq:zerothOrderConstraintPalatini}
  0 = \frac{\delta\phi}{\alpha}
\end{equation}
meaning that $\delta\phi$ must be exactly zero, which means you in this case are not allowed to deviate from GR. This means that the theory must in fact be an exact $f(R)$ theory for $\mathcal{C}$ to be close to constant in this scheme. Going back to the assumptions above, it is in fact enough that they take the form described above to first order for this to hold true. To be completely precise the expression (\ref{eq:zerothOrderConstraintPalatini}) means that $\delta\phi$ is zero to first order in $\alpha$ so $\delta\phi$ could be a second order perturbation. We must then go to first order in $\alpha$ in equation (\ref{eq:NumSolveForddotphiCoeffs}) to get an expression for its value. Here all the other $\delta\phi$ and derivatives of it disappear as they are second order small, and the equation (\ref{eq:NumSolveForddotphiCoeffs}) becomes an expression for $\delta\phi$. We can do a double derivative of that expression and compare to the second order expansion of equation (\ref{eq:NumSolveForddotphiCoeffs}), and the results will again not be identical unless some constraints on the matter content is satisfied as seen in the previous section hold. 

\subsection{The difference between C-theory and $f(R)$}
The question now is, what happens if we assume $\mathcal{C}$ to be exactly constant, that is we revert to a normal $f(R)$ theory? The assumptions then become:\footnote{Multiplying $\mathcal{C}$ by a constant can in this case be put into $f$ and/or a redefinition of the Planck mass therefore we simply assume it to have unit value.}
\begin{eqnarray}
  \phi = \mathcal{R} = R + \delta\phi \qquad \mathcal{C} = 1\nonumber\\
  f(R) = R + \delta f
\end{eqnarray}
Again we take $R$ to be the exact GR response to the underlaying universe components, rather than the Ricci tensor of the geometry. We then essentially have two cases. Either $f''(R) = 0$ or it is not.\footnote{The first case can be exchanged with $f''(R)$ being 0 to second order.} In the first case $A$ is exactly zero and it makes no sense to divide by it. This is because the expression we had for $\ddot{\phi}$ actually has no $\ddot{\phi}$ term, and the equations from which they stem yield equations only for $\dot{\phi}$ and $\phi$. Assuming that $f'''$ is also zero the resulting equation just becomes:
\begin{eqnarray}
  0 = \delta\phi + 2\delta f - \left(R + \delta\phi\right)\delta f'
\end{eqnarray}
To first order this just means that $\delta \phi$ makes up for whatever constant shift $f$ may have from being a linear function in $R$, which is fine.

If $\delta f''$ and $\delta f'''$ are first order perturbations, the equation comes out as:
\begin{equation}
  \ddot{R} + \delta\ddot{\phi} = \frac{\delta f'''}{\delta f''}\left(\dot{R}^2 + 2\dot{R}\delta\dot{\phi}\right) + \left(\frac{R + \delta\phi}{6H^2} + 1\right)\left(\dot{R} + \delta\dot{\phi}\right) + \frac{1}{3H^2\delta f''}\left(2\delta f - R\delta f' + \delta\phi - \delta f'\delta\phi\right) 
\end{equation}
to first order. Separating the zeroth and first equations they become respectively:
\begin{eqnarray}
  \ddot{R} &=& \frac{\delta f'''}{\delta f''}\ddot{R} + \left(\frac{R}{6H^2} + 1\right)\dot{R} + \frac{2\delta f - R\delta f' + \delta\phi}{3H^2\delta f''}\nonumber\\
  \delta\ddot{\phi} &=& 2\frac{\delta f'''}{\delta f''}\dot{R}\delta\dot{\phi} + \dot{R}\frac{\delta\dot{\phi}}{6H^2} + \left(\frac{R}{6H^2} + 1\right)\delta\dot{\phi} + \frac{\delta f'\delta\phi}{3H^2\delta f''}
\end{eqnarray}
Here we again see that the zeroth order equation shows how the change in $\delta\phi$ must be there to make up for the change in $f$, $\delta f$. Therefore we do not get $\delta\phi$ only in terms of the GR quantities, and the theory does not give constraints for the universe content as was seen in the cases where $\mathcal{C}$ was not constant.

 \section{Conclusions}\label{sec:Conclusions}
 In this note we presented the general field equations of C-theories, as well as the FRW equations. We gave a reformulation of these which could be applicable to numerics and linearisation. When looking at linearised theories close to GR we found that many of these are unstable or lead to inconsistencies like constraints on the possible universe components. Though these findings are not completely generic we found that they will hold for quite a wide range of theories close enough to GR. Whether this generalises further should be a subject for further study.

Though this seems to put strong constraints on the applicability of the theories, there might still be opportunities for these theories formulated seemingly much further from their GR limits. Whatever the result of further generalised studies, this makes numerical work on what may remain of viable theories difficult as making appropriate choices for at the same time stable and physically meaningful initial conditions may be challenging.\footnote{Note, however, that this is only important to this theory if applied as a late time component to cosmology. If the interest is focused towards inflation, the relative stability close to GR is not necessary.} For the theories already discussed in their close to GR limit, this may even be impossible.

Because of these inconsistencies close to GR, we have not made a numerical study of the cosmological evolution of this theory. Such a study was done in \cite{Finke2012}, with results that appear to qualitatively agree with our conclusions.

%however, it is unclear to us how the initial conditions were set in that case.

\begin{acknowledgments}
D.F.M. thanks the Research Council of Norway FRINAT grant 197251/V30.
D.F.M. is also partially supported by project CERN/FP/123618/2011 and CERN/FP/123615/2011. TK is supported by the Research Council of Norway.
\end{acknowledgments}

  %bibliography:
  \bibliographystyle{JHEP}
  \bibliography{sources}

\end{document}